\documentclass[aps,pra,twocolumn,showpacs,superscriptaddress,groupedaddress]{revtex4}

\usepackage{amsmath, hyperref, braket, amssymb, natbib, graphicx,float}

\newcommand{\diff}{\ensuremath{\mathrm{d}}}
\newcommand{\EXP}[1]{\ensuremath{\mathrm{e}^{#1}}}

\newcommand{\dB}{\ensuremath{~\mathrm{dB}}}
\newcommand{\Hz}{\ensuremath{~\mathrm{Hz}}}
\newcommand{\MHz}{\ensuremath{~\mathrm{MHz}}}

\usepackage[dvipsnames]{color}

\newcommand{\refsec}[1]{Sec.~\ref{#1}}
\newcommand{\refeq}[1]{Eq.~(\ref{#1})}
\newcommand{\reffig}[1]{Fig.~\ref{#1}}
\newcommand{\reftab}[1]{Table~\ref{#1}}

\begin{document}

\title{Arbitrary multimode Gaussian operations on mechanical cluster states}
\author{Darren W. Moore}
\email{dmoore32@qub.ac.uk}
\affiliation{School of Mathematics and Physics, Queen's University Belfast, BT7 1NN, UK}
\author{Oussama Houhou}
\email{o.houhou@qub.ac.uk}
\affiliation{School of Mathematics and Physics, Queen's University Belfast, BT7 1NN, UK}
\affiliation{Laboratory of Physics of Experimental Techniques and Applications, University of Medea, Medea 26000, Algeria}
\author{Alessandro Ferraro}
\email{a.ferraro@qub.ac.uk}
\affiliation{School of Mathematics and Physics, Queen's University Belfast, BT7 1NN, UK}

\begin{abstract}

We consider opto- and electro-mechanical quantum systems composed of a driven cavity mode interacting with a set of mechanical resonators. It has been proposed that the latter can be initialized in arbitrary cluster states, including universal resource states for Measurement Based Quantum Computation (MBQC). We show that, despite the unavailability in this set-up of direct measurements over the mechanical resonators, computation can still be performed to a high degree of accuracy. In particular, it is possible to indirectly implement the measurements necessary for arbitrary Gaussian MBQC by properly coupling the mechanical resonators to the cavity field and continuously monitoring the leakage of the latter. We provide a thorough theoretical analysis 
of the performances obtained via indirect measurements, comparing them with what is achievable when direct measurements are instead available. We show that high levels of fidelity are attainable in parameter regimes within reach of present experimental capabilities.
\end{abstract}

\maketitle

\section{Introduction}
\label{intro}

One of the major expected outcomes of research into quantum technologies is the production of a quantum computer, a device which allows efficient solution of problems considered inefficient classically \cite{nielsen2010quantum,ladd2010quantum}. Amongst the various emerging platforms for quantum technologies is that of quantum optomechanics, in which radiation pressure is exploited to establish a quantum dynamics between mechanical and radiative systems \cite{milburn2011optomechrev,meystre2013short,aspelmeyer2014cavity}. 
This radiation pressure coupling finds expression in a large range of settings, from small micromechanical resonators  \cite{ding2011wavelength,park2009resolved,schliesser2008resolved,verhagen2012quantum,schliesser2009resolved} to larger mesoscopic systems \cite{Connell:2010,sekatski2014macroscopic,mancini1998optomechanical,chen2013macroscopic,purdy2013observation,tan2013generation,abdi2016dissipative}, electromechanical systems \cite{teufel:2011,wollman2015quantum,weinstein2014observation,pirkkalainen2015squeezing} and more recently in systems of levitated particles \cite{genoni2015quantum, bhattacherjee2016optomechanical, ortiz2017continuous}. Along with this plethora of technical settings comes an abundance of applications for quantum technologies, including hybrid quantum information processing \cite{rogers2014hybrid}, cooling of macroscopic objects to the ground state \cite{Connell:2010,teufel:2011,chan:2011,safavi:2012,brahms2012optical,verhagen2012quantum}, back action evading measurements \cite{clerk2008back, hertzberg2010back, woolley2013two} and preparation of non-classical states \cite{abdi2016dissipative, rips2012steady, rips2014nonlinear, khosla2013quantum, vanner2011selective, sankey2010strong}.

Additionally, recent experiments demonstrated the possibility to coherently couple multiple mechanical resonators to a single cavity field \cite{massel2012multimode, ockeloen2016quantum, damskagg2016dynamically, grass2016optical}. In fact, various theoretical analyses of multiple resonators coupled to radiation pressure have been put forward \cite{Bhattacharya2008, hartmann2008steady, chang2011slowing, stannigel2012optomechanical, schmidt2012optomechanical, xuereb2012strong, tomadin2012reservoir, ludwig2013quantum, seok2013multimode, xuereb2014reconfigurable, nielsen2016multimode, li2017enhanced}. In particular, a recent proposal showed that arbitrary graph states may be generated in an array of resonators immersed in a cavity field \cite{houhou2015generation} using a generalisation of the reservoir engineering sometimes used for cooling \cite{wollman2015quantum,lei2016quantum,pirkkalainen2015squeezing,massel2012multimode}. As well, a method for reconstructing the state of a network of harmonically interacting resonators coupled to radiation pressure has been proposed \cite{moore2016Recon}. This provides a promising opportunity to use optomechanics as a platform for Measurement Based Quantum Computation (MBQC) \cite{Raussendorf:01,menicucci2006universal} in the continuous-variable setting \cite{Ferraro05,Weedbrook:12,adesso2014continuous}. The main advantage would be that, being hosted in stationary or solid-state based architectures, they offer a promising path towards integrated and scalable quantum technologies. However, there is an inherent obstacle that could potentially frustrate this opportunity. In fact, in the typical scenario of MBQC, the measurements are projective and performed directly on the nodes of the cluster \cite{Weedbrook:12}. However, in the side-band resolved regime considered in Ref.~\cite{houhou2015generation,moore2016Recon}, mechanical modes are inaccessible to direct measurement. Thus, it is necessary to devise indirect measurement strategies which unfortunately typically introduce noise in the process. The latter could represent a significant hindrance to computation, spoiling the operational performances of MBQC. In other words, in order to fully exploit quantum opto-mechanical systems for advanced computational purposes, it is necessary to identify an effective indirect-measurement strategy and assess its performances in detail. 

Here we address this issue by proposing a method for implementing arbitrary single- and multi-mode Gaussian operations on the mechanical cluster state, using continuous monitoring \cite{jacobs2006straightforward, genoni2016conditional,doherty1999feedback,genoni2015quantum} of an observable coupled to the nodes of the cluster. 
This is accomplished using a quantum non-demolition (QND) interaction of the cavity field with the mechanical cluster node to be measured \cite{Braginsky547,braginsky1996quantum}. This interaction drives the latter towards an eigenstate of the chosen observable. We must ensure that the result is in accord with the usual procedure of MBQC, which as said is based on direct measurements of the cluster nodes. To this aim, we provide a thorough theoretical analysis of the performances obtained via indirect measurements, comparing them with what is achievable when direct measurements are instead available \cite{Weedbrook:12}, as in the more common purely optical scenario \cite{chen2014experimental, roslund2014wavelength, yoshikawa2016invited}. The dependence on the system parameters is analysed in detail, by identifying the ones that most affect the performances of a set of universal Gaussian operations. This study can help in devising experimental settings with the purpose of implementing MBQC over mechanical resonators. In particular, we show that high levels of fidelity, compared with direct measurements, are attainable in parameter regimes within reach of present experimental capabilities.

The article is arranged as follows: in \refsec{sec:MBQC} we discuss the relevant theory of MBQC with continuous variables (CV) and in \refsec{sec:gaussian-conditional-dynamics} we discuss the technique of continuous monitoring. We included these two sections for completeness and to set the notation, however the reader familiar with both topics can safely move to \refsec{sec:optomechanical-setting}, where we discuss the optomechanical setup we envision and how to apply continuous monitoring to it. Then, in \refsec{sec:continuous-monitoring} we demonstrate that continuous monitoring successfully reproduces the results of standard MBQC using projective measurements directly onto the cluster state and we analyze in detail the effects that various types of noise parameters have on the performance of MBQC. Finally, the conclusion is given in \refsec{sec:conclusion}.

\section{Measurement Based Quantum Computation over continuous variables}\label{sec:MBQC}

In the common circuit model of quantum computation, the system is initialized in a blank register and the computation proceeds via single mode and entangling unitary gates \cite{nielsen2010quantum,lloyd2003quantum}. MBQC is an alternate model of computation that instead uses local measurements on a highly entangled resource state to drive the computation \cite{Raussendorf:01, menicucci2006universal}. The resource state is modelled by a graph $\mathbf{G}=\{V,E\}$, where $V$ and $E$ are the sets of vertices and edges respectively. Physically, the vertices are represented by states embodying a balanced superposition of the states of the computational basis. For example in the case of qubits these are the states $\ket{+}=\frac{\ket{0}+\ket{1}}{\sqrt{2}}$. For CV they are a collection of zero momentum eigenstates $\ket{0}_p=\int\diff x\ket{x}_q$ typically approximated by highly squeezed vacuum states (where $\ket{x}_q$ represents the position eigenstate with eigenvalue $x$). The edges of the graph are implemented via (entangling) control-phase operations on pairs of vertices, which for CV are simply represented by $CZ_{jk}=\EXP{iq_jq_k}$, where $j,k$ denotes the vertices under consideration and $q_j$ is the position operator of mode $j$. A graph state suitable for universal computation is called a cluster state and consists of a two-dimensional grid \cite{menicucci2006universal,zhang2006continuous}.

\begin{figure}
\includegraphics[width=.6\columnwidth]{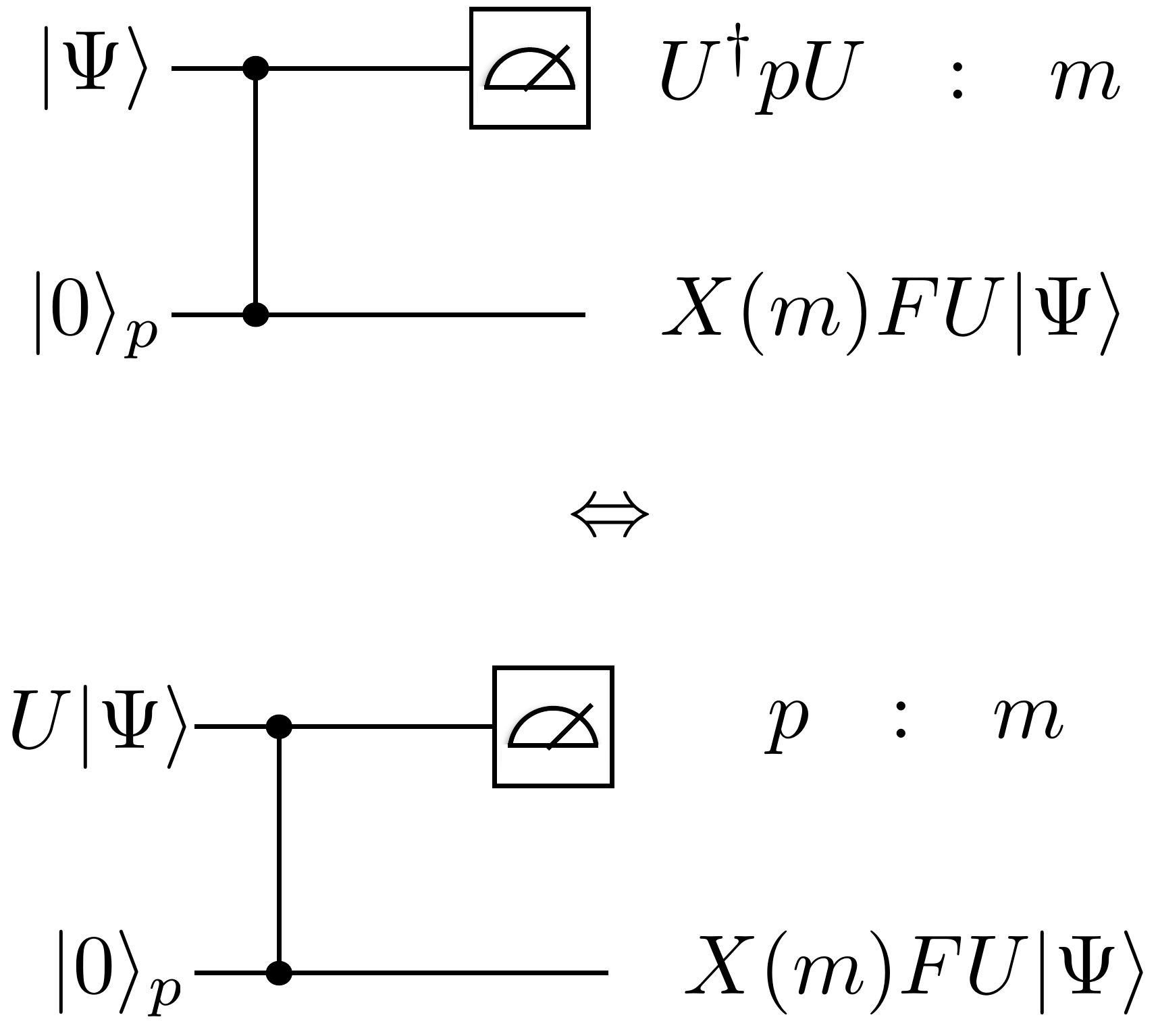}
\caption{The circuit representing gate teleportation. (Upper circuit) An input state $\ket{\Psi}$ is linked via a control phase gate to an ancilla state $\ket{0}_p$. A measurement of $p$ with outcome $m$ teleports the input state to the ancilla along with some by-product known operations ($F$ being the Fourier transform and $X(m)=e^{-imp}$). (Lower circuit) If a unitary operation $U$ is carried out on the input before the entangling gate then it also will be teleported, as $U\ket{\Psi}$ is simply another legitimate input state. However, if $U$ is diagonal in the computational basis then the two circuits are equivalent, since $U$ commutes with the control-phase operation and can be incorporated into the measurement by a rotation of the measurement basis. Thus the measurement itself can be used to induce the operation $U$ on an input state.}
\label{Cirque}
\end{figure}

Operations on the cluster are performed by local measurements on the nodes which drive the rest of the cluster into a new state. This is most clearly seen from gate teleportation (\reffig{Cirque}), whereby a measurement on some input state induces the action of a unitary gate on a copy of the input teleported into an ancilla state. A series of such measurements chosen appropriately will drive the remainder of the cluster into a state representing the output of a computation. Each measurement step will accumulate known by-product operations that can be corrected: phase space displacements $X(m)=e^{-imp}$, which depend on the measurement outcome $m$, and Fourier operations. The cluster state is composed of Gaussian states, meaning that it is characterised completely by its first and second moments $\mathbf{d}=\braket{\hat{\mathbf{r}}}$ and $\sigma_{ij}=\frac{1}{2}\braket{r_ir_j-r_jr_i}-\braket{r_i}\braket{r_j}$ where $\hat{\mathbf{r}}=\begin{pmatrix}q_1, & p_1, & \dots, q_n, & p_n\end{pmatrix}^\top$, $q_j=\frac{a_j+a_j^\dagger}{\sqrt{2}}$ and $p_j=\frac{i(a_j^\dagger-a_j)}{\sqrt{2}}$, and $a_j$ is the annihilation operator for a bosonic mode describing the $j$-th node of a cluster. For Gaussian operations on Gaussian cluster states, the local displacements can be discarded from the analysis, since they may be applied at any point in the computation. This leaves the output with a Fourier transform still applied. The protocol of gate teleportation shows that to apply a unitary operation $U$ via measurements, one should measure in the quadrature basis $U^\dagger pU$.

In order to achieve universal computation (with CV), a certain minimal set of operations are required \cite{lloyd98computation,menicucci2006universal}. These include a universal set of single-mode Gaussian transformations: arbitrary phase-space displacements $D(\alpha)=\EXP{\alpha a^\dagger-\alpha^*a}$, the Fourier gate $F=\EXP{\frac{i\pi}{2}}\EXP{i\pi a^\dagger a}$ and the shearing gate $S(\lambda)=\EXP{i\lambda q^2}$. Adding a single multimode Gaussian operation suffices to cover all Gaussian transformations. A typical choice for such an operation is the control-phase gate already considered in the procedure to generate the cluster \cite{phasegate}. Finally, at least one non-Gaussian element must also be included. Typically this is an operation of the form $\EXP{itq^n}$ for some $t$ and $n\geq 3$.

\begin{figure}
\includegraphics[width=\columnwidth]{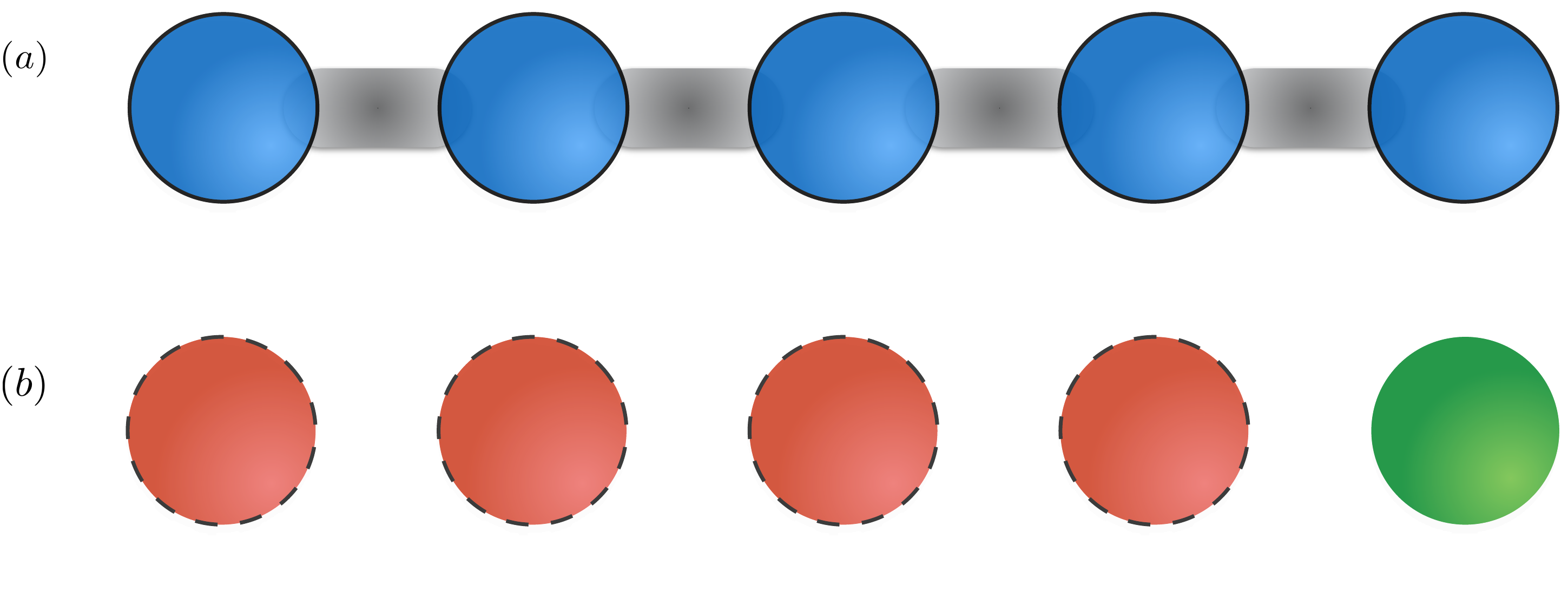}
\includegraphics[width=\columnwidth]{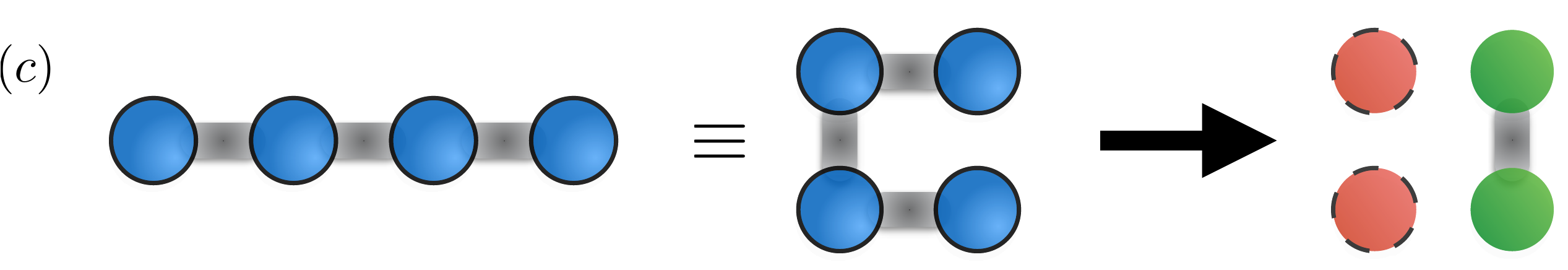}
\caption{(colour online) A linear cluster of five nodes. Before measurement commences (a) all nodes (blue/solid border) are equally linked by balanced CZ gates and each node has an equal degree of squeezing. After a sequence of measurements (b) the (red/dashed border) measured nodes are disconnected from the cluster and are discarded, leaving behind a single node (green/no border) modified by the required operation. A two-mode operation requiring a dual-rail can be minimally simulated by (c) taking the four-node ancilla and manipulating it topologically.}
\label{Cluster}
\end{figure}

Given the availability of quadrature measurements (homodyne detection), at the level of state covariances any single-mode Gaussian transformation can be achieved using the shearing gate and Fourier transform. In other words, a single step gate teleportation is encapsulated by $FS(\lambda)$, for a particular value of $\lambda$. The measurement associated with the shearing gate is $p+\lambda q$. With an ancillary linear cluster of just four nodes the full range of single mode transformations can be applied to some input state, including the corrective Fourier gates \cite{ukai2010universal,ukai2011demonstration}. Each single-mode Gaussian unitary can be decomposed into a $2\times 2$ symplectic matrix $\mathcal{M}$. For example, the Fourier transform and the shearing gate are associated with the matrices $f=\begin{pmatrix}0 & -1\\1 & 0\end{pmatrix}$ and $s(\lambda)=\begin{pmatrix}1 & 0\\ \lambda & 1\end{pmatrix}$ respectively. In particular, given a sequence of four gate teleportations defined by $F S(\lambda_4)F S(\lambda_3)F S(\lambda_2)F S(\lambda_1)$ the symplectic transformation associated to it is given by
\begin{align}
\label{symplectic_decomposition}
&\mathcal{M}=\nonumber\\ &\begin{pmatrix}
\lambda_4\lambda_3(\lambda_2\lambda_1-1)-\lambda_1(\lambda_2+\lambda_4)+1 
& \lambda_4\lambda_3\lambda_2-\lambda_4-\lambda_2
\\
-\lambda_3\lambda_2\lambda_1+\lambda_3+\lambda_1 & -\lambda_3\lambda_2+1
\end{pmatrix}\ .
\end{align}
The ability to tune each parameter $\lambda_j$ above enables one to implement an arbitrary $2\times 2$ symplectic matrix. Thus, by applying a sequence of four teleportations (with tunable $\lambda_j$) to an initial five-mode cluster, any single-mode Gaussian transformation can be implemented (\reffig{Cluster}). We will use this formalism to demonstrate universal single-mode operations. In particular, we will use the notation $M \rightarrow \{\lambda_1,\lambda_2,\lambda_3,\lambda_4\}$ to associate with a given single-mode operation $M$ the corresponding set of four measurements $p+\lambda_jq$ that implement it ($j=1,\dots,4$).

For the CZ gate, one requires at least two linear graphs linked into a grid. The simplest version of this is just the same four node linear cluster re-arranged so that the first two nodes in the dual-rail are the middle nodes of the linear cluster. To provide a multimode operation, we use a technique for shaping cluster states called wire shortening \cite{menicucci2006universal,zhang2006continuous}. It so happens that a measurement of $p$ on a cluster node deletes the vertex while maintaining the edges to which it was connected. A succession of two such measurements on the leftmost column of the right hand side of \reffig{Cluster}~(c) deletes the vertices in such a way that the remainder of the cluster is given by the two end nodes (right column) connected by an edge. Creating this edge is equivalent to performing a CZ operation between the two end nodes.

\section{Gaussian Conditional 
Dynamics}\label{sec:gaussian-conditional-dynamics}

Let us briefly review the formalism needed to describe Gaussian 
conditional dynamics via continuous monitoring. Consider a set of $m$ 
input modes interacting with $n$ system modes, where the former are 
associated with Markovian bath modes and the latter with 
the system of interest. As said, in the Gaussian regime, the dynamics can 
be described by the first and second moments of the canonical operators. 
The coupling between the modes is at most quadratic,
\begin{equation}
     \hat{H}_C=\frac{1}{2}\hat{\mathbf{r}}^\top_{\text{SB}}
     \begin{pmatrix}
         0 & C\\
         C^\top & 0
     \end{pmatrix}
\hat{\mathbf{r}}_{\text{SB}}\equiv\frac{1}{2}\hat{\mathbf{r}}^\top_{\text{SB}}H_C\hat{\mathbf{r}}_{\text{SB}}\ 
,
\end{equation}
where $C$ is a $n\times m$ real matrix characterising the interaction 
between system (S) and input (B) modes, and we use the quadrature ordering 
$\hat{\mathbf{r}}_{\text{SB}}^\top\equiv\begin{pmatrix}\hat{\mathbf{r}}^\top, 
& 
\hat{\mathbf{r}}_{\text{in}}^\top\end{pmatrix}\equiv\begin{pmatrix}q_1,p_1,\dots,q_n,p_n,Q_1,P_1,\dots,Q_m,P_m\end{pmatrix}^\top$ 
with system and input modes denoted by lowercase and uppercase symbols 
respectively. The initial covariance matrix $\sigma\oplus\sigma_B$, 
where $\sigma$ is the initial state of the system and $\sigma_B$ is the 
initial state of the input modes, evolves according to the symplectic 
transformation
\begin{equation}
     \EXP{\Omega H_C\diff w}=\mathbb{I}+\Omega H_C\diff w+\frac{(\Omega 
H_C)^2}{2}\diff t+o(\diff t)\ ,
\end{equation}
where the Wiener process $\diff w$ satisfies the Ito rule $\diff 
w^2=\diff t$ \cite{genoni2016conditional}. The symplectic form $\Omega$ 
is defined as
\begin{equation}
\Omega=\bigoplus_{j=1}^k\omega\qquad;\qquad\omega=\begin{pmatrix}0 & 1 
\\ -1 & 0 \end{pmatrix}\ .
\end{equation}
For brevity, we do not indicate explicitly the dimension of $\Omega$ which should be 
extracted from the context. The evolution of system-input modes is seen to be  
\cite{genoni2016conditional}
\begin{align}
&\EXP{\Omega H_C\diff w}(\sigma\oplus\sigma_B)\EXP{\Omega H^\top_C\diff 
w}\nonumber\\=
&(\sigma\oplus\sigma_B)+\Big(\frac{\Omega C\Omega C^\top\sigma+\sigma 
C\Omega C^\top\Omega}{2}\Big)\oplus\hat{\sigma}_{B,1}\diff t\nonumber\\
&+\Omega 
C\sigma_BC^\top\sigma_BC^\top\Omega^\top\oplus\hat{\sigma}_{B,2}\diff 
t+\sigma_{SB}\diff w+\sigma_{BB}o(\diff t)\ ,
\end{align}
with
\begin{align}
&\sigma_{SB}=\begin{pmatrix}0 & \Omega C\sigma_B+\sigma C\Omega^\top\\
\sigma_BC^\top\Omega^\top+\Omega C^\top\sigma & 0
\end{pmatrix}\ ,\\
&\hat{\sigma}_{B,1}=\frac{\Omega C^\top\Omega 
C\sigma_B+\sigma_BC^\top\Omega C\Omega}{2}\ ,\\
&\hat{\sigma}_{B,2}=\Omega^\top C^\top\sigma C\Omega\ .
\end{align}

The behaviour of the system when all the input modes of the bath are disregarded is then 
encapsulated in the matrix diffusion equation, or Lyapunov equation,
\begin{equation}\label{lyapunov}
     \dot{\sigma}=A\sigma+\sigma A^\top+D\ ,
\end{equation}
with drift and diffusion matrices
\begin{equation}
     A=\frac{\Omega C\Omega C^\top}{2}\qquad;\qquad D=\Omega 
C\sigma_BC^\top\Omega^\top\ .\label{driftdiffusion}
\end{equation}
Interaction among the system modes leads to an additional Hamiltonian 
term, $\hat{H}_s=\frac{1}{2}\hat{\mathbf{r}}^\top H_s\hat{\mathbf{r}}$, 
which modifies the drift matrix:
\begin{equation}
     A\rightarrow A+\Omega H_s\ .\label{driftsystem}
\end{equation}
This interaction term is crucial, as it will contain the QND interaction 
that allows the measurement of mechanical quadratures.

So far this formalism describes noise. Let us assume now that the input modes undergo continuous monitoring instead. The description of Gaussian measurements on Gaussian states involves an instantaneous 
mapping of the covariance matrix and the vector of first moments \footnote{
The latter are not relevant for the purpose of this paper, since for Gaussian MBQC we can always absorb first moment displacements in by-product operations}. Denoting the post-measurement covariance matrix of the measured modes of the bath B by $\sigma_m$, the post-measurement covariances of the system modes are 
given by
\begin{equation}
\sigma\rightarrow\sigma-\sigma_{C}\frac{1}{\sigma_B+\sigma_m}\sigma_{C}^\top\ 
.
\end{equation}
Here $\sigma_C$ is the off-diagonal block of $\sigma_{SB}$ . Applying this 
results in a Riccati equation (rather then in a Lyapunov equation as in \refeq{lyapunov}):
\begin{equation}
\label{riccati}
\dot{\sigma}=\tilde{A}\sigma+\sigma\tilde{A}^\top+\tilde{D}-\sigma 
BB^\top\sigma\ ,
\end{equation}
where now we defined
\begin{align}
     &\tilde{A}=A-\Omega C\sigma_B\frac{1}{\sigma_B+\sigma_m}\Omega 
C^\top\ ,\label{Rdriftdiffusion}\\
     &\tilde{D}=D+\Omega 
C\sigma_B\frac{1}{\sigma_B+\sigma_m}\sigma_BC^\top\Omega\ ,\\
     &B=C\Omega\sqrt{\frac{1}{\sigma_B+\sigma_m}}\ .\label{Rdriftdiffusion1}
\end{align}

In order to model the opto-mechanical system considered for MBQC, we need to consider a situation in which only a portion of the input modes undergo monitoring, whereas the rest are lost as genuine noise. To take account of this, a small modification is required. The modes undergoing purely dissipative dynamics entail an evolution of the system covariance matrix in Lyapunov form of \refeq{lyapunov}, whereas those undergoing monitoring evolve the covariance matrix in the Riccati form of \refeq{riccati}. We introduce a distinction between the interaction of the monitored and dissipative modes with the system modes via $C_m$ and $C_d$. $C_d$ enforces a Lyapunov equation with matrix coefficients $A_\text{Lyap}$ and $D_\text{Lyap}$ whereas $C_m$ enforces a Riccati equation with coefficients $\tilde{A}_\text{Ricc}$, $\tilde{D}_\text{Ricc}$ and $B$ that depends only on $C_m$. The linearity of these equations allows them to be simply added to find the full dynamics of the covariance matrix under both effects
\begin{align}
\dot{\sigma}=&(A_\text{Lyap}+\tilde{A}_\text{Ricc})\sigma+\sigma(A_\text{Lyap}^\top+\tilde{A}_\text{Ricc}^\top)\nonumber\\&+D_\text{Lyap}+\tilde{D}_\text{Ricc}-\sigma BB^\top\sigma\,. \label{LyapRicc}
\end{align}

Finally for homodyne detection, inefficient detectors can be introduced 
by distorting the post-measurement state as follows
\begin{equation}
\sigma_m\rightarrow\frac{1}{\eta}\sigma_m+\frac{1-\eta}{\eta}\mathbb{I}\ .
\end{equation}

In the following we are going to consider a homodyne detection scheme. The post-measurement state of the homodyned mode will be a position-squeezed vacuum with squeezing parameter denoted as $r_\text{post-meas}$, hence $\sigma_m=\rm{\frac{1}{2}Diag(e^{-2r_\text{post-meas}},e^{2r_\text{post-meas}})}$. Notice that perfect homodyne detection requires $r_\text{post-meas}\rightarrow\infty$. Note that, in contrast to optical schemes in which a mode representing a cluster vertex is destroyed by the measurement, the measured mechanical mode is disconnected from the cluster by the measurement in the sense of having its correlations with the remainder of the cluster destroyed by the measurement.

\section{Optomechanical Setting and Mechanical Cluster 
States}\label{sec:optomechanical-setting}

As mentioned in \refsec{intro}, the mechanical portion of an 
optomechanics experiment is inaccessible to direct measurement in the 
regimes of our interest. However, schemes exist for the precise 
measurement of the mechanical quadratures via measurements on the cavity 
field \cite{clerk2008back, hertzberg2010back, woolley2013two}. One of these involves continuous monitoring by engineering a QND interaction 
between the cavity field and the mechanical quadrature operator one is 
interested in measuring. We are now going to show how this interaction 
can be exploited for our purposes.

\begin{figure}
     \includegraphics[width=\columnwidth]{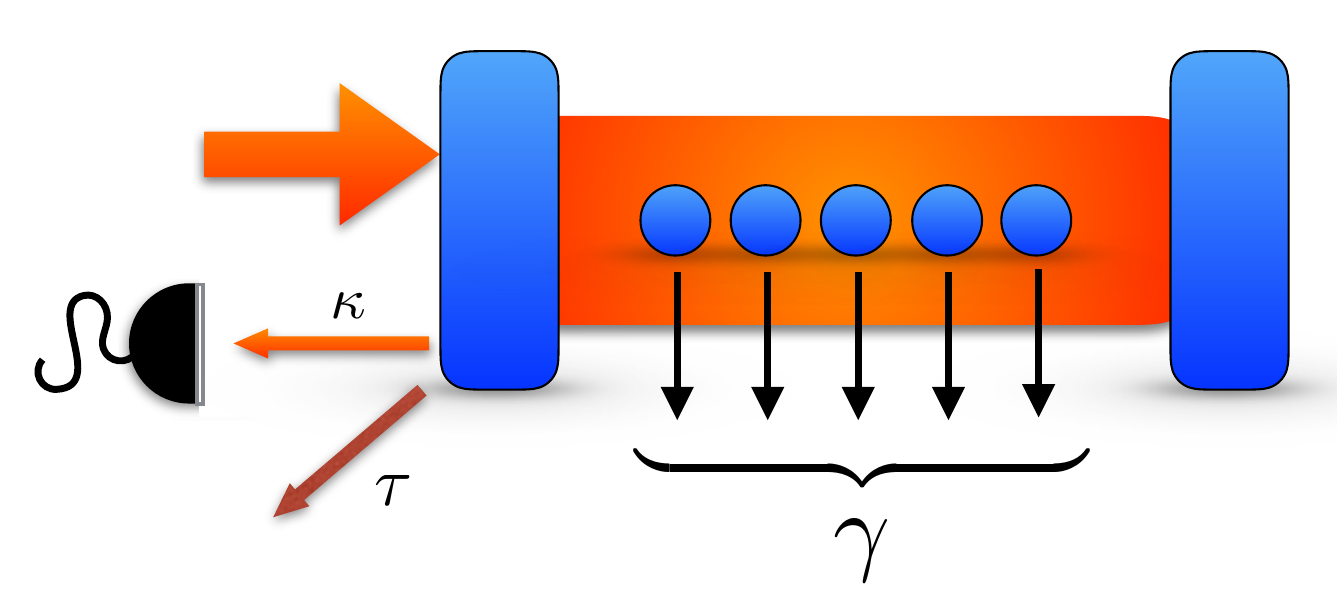}
     \caption{A driven single-mode cavity hosts a collection of 
non-interacting mechanical resonators. The latter are supposed to be 
already prepared in a cluster state and to have non-overlapping 
frequencies (this can be achieved following the proposal of 
Ref.~\cite{houhou2015generation}). They are coupled to a thermal 
environment with dissipation rate $\gamma$. The cavity decays at a rate 
$\kappa$ into a monitored mode which is eventually measured via a 
homodyne detector. Unmonitored losses at rate $\tau$ are also taken into 
account.}
     \label{optomech}
\end{figure}

The system we focus on is pictorially represented in \reffig{optomech}. 
It is composed of an array of $N$ non-interacting resonators immersed in 
the cavity field which dissipates at a rate $\kappa$. The resonators are 
assumed to have identical mechanical damping rates $\gamma$ which 
dissipate into thermal baths of temperature $T$. The cavity operates in 
the sideband resolved regime $\kappa\ll\text{min}(\Omega_j)$ with 
$\Omega_j$ the mechanical frequencies, and is driven by a collection of 
$M$ fields. The collection of resonators are supposed to have 
non-overlapping frequencies, allowing us to address each of them 
individually for measurement.

The linearised Hamiltonian for the described system, in an interaction 
picture with respect to the free evolution, takes the form
\begin{equation}
H=a^\dagger\sum_j^N\sum_k^M\alpha_k\EXP{i\phi_k}g_j\EXP{i(\omega_c-\omega_k)t}(b_j\EXP{-i\Omega_jt}+b_j^\dagger 
\EXP{i\Omega_jt})+\text{h.c.}\ ,
\end{equation}
where $g_j$ are the single photon-phonon interaction strengths, 
$\Omega_j$, $\omega_c$ and $\omega_k$ are the frequencies of the 
resonators, cavity and drives respectively, $\alpha_k\geq0$ are the 
cavity drive strengths and $\phi_k$ their phases, whereas $a$ and $b_j$ 
are the annihilation operators for the cavity and mechanical modes. We 
will assume that $\gamma\ll\kappa$. In addition, we will work in the side-band resolved regime $\kappa\ll\text{min}(\Omega_j)$ and assume weak coupling between radiation and mechanics:
$g_j\alpha_j\ll\text{min}(\Omega_j)$.

Now choose two drive frequencies for each resonator $j$ to be on the 
mechanical sidebands $\omega_j^\pm=\omega_c\pm\Omega_j$ with strengths 
$\alpha_j^\pm$ and phases $\phi_j^\pm$:
\begin{align}
H&=a^\dagger\sum_j^{N}g_j\left[\alpha_j^+\EXP{i\phi_j^+}\EXP{-i\Omega_j t}\left(b_j\EXP{-i\Omega_jt}+b_j^\dagger 
\EXP{i\Omega_jt}\right)+\right.\nonumber\\
&\left.\alpha_j^-\EXP{i\phi_j^-}\EXP{i\Omega_j t}\left(b_j\EXP{-i\Omega_jt}+b_j^\dagger 
\EXP{i\Omega_jt}\right)\right]+\text{h.c.}\nonumber\\
&=a^\dagger\sum_jg_j\left[\alpha_j^+\EXP{i\phi_j^+}\left(b_j^\dagger+b_j\EXP{-2i\Omega_jt}\right)\right.\nonumber\\
     &\left.+\alpha_j^-\EXP{i\phi_j^-}\left(b_j^\dagger 
\EXP{2i\Omega_jt}+b_j\right)\right]+\text{h.c.}\ .
\end{align}
Let $\alpha_j=\alpha_j^+=\alpha_j^-$ and $\phi_j=\phi_j^+=-\phi^-_j$. Then
\begin{align}
H&=(a+a^\dagger)\sum_jg_j\alpha_j\left[\EXP{i\phi_j}\left(b_j^\dagger+b_j\EXP{-2i\Omega_jt}\right)\right.\nonumber\\
     &\left.+\EXP{-i\phi_j}\left(b_j+b_j^\dagger 
\EXP{2i\Omega_jt}\right)\right]\nonumber\\
&=\sqrt{2}(a+a^\dagger)\sum_jg_j\alpha_j\left[X_{\phi_j}+\right.\nonumber\\
&\left.\frac{1}{\sqrt{2}}\left(b_j\EXP{-i(2\Omega_jt-\phi_j)}+b_j^\dagger 
\EXP{i(2\Omega_jt-\phi_j)}\right)\right]\nonumber\\
&=\sqrt{2}(a+a^\dagger)\sum_jg_j\alpha_j\left[X_{\phi_j}+X_j\cos(2\Omega_jt-\phi_j)\right.\nonumber\\
     &\left.+P_j\sin(2\Omega_jt-\phi_j)\right]\ ,
\end{align}
where we have defined the quadratures for the mechanical resonators as follows:
\begin{align}\label{X_phi}
X_{\phi_j} &=X_j\cos\phi_j+P_j\sin\phi_j \;,
\end{align}
with $X_j =\frac{b_j+b^\dagger_j}{\sqrt2}$ and $P_j =\frac{b_j-b^\dagger_j}{i \sqrt2}$.
Since we are in the weak coupling regime we may take the rotating wave approximation, discarding the fast rotating terms so that this Hamiltonian is a time averaged interaction of the cavity position with an arbitrary 
quadrature of each mechanical mode. Since the interaction with each 
resonator depends on the choice of $\alpha_j$ one can imagine a step by 
step process in which each resonator is addressed in turn by having all 
but one of the $\alpha_j$ set to zero. In such a scenario, the 
Hamiltonian during one step is, in the above time-averaged sense, a QND 
interaction of cavity-position $X=\frac{a+a^\dagger}{\sqrt{2}}$ with 
$X_\phi$ of a single resonator, say the $k$-th one:
\begin{equation}
     H_k=2g_j\alpha_k XX_{\phi_k}\ .
\end{equation}

Continuous measurement of the position quadrature via homodyne detection 
on the output field from the cavity produces a back action evading 
measurement of $X_{\phi_k}$. This provides an effective indirect 
measurement of the mechanical resonator which is in fact driven towards 
a highly squeezed approximation to an eigenstate of $X_{\phi_k}$. In 
turn, the remainder of the cluster is driven to the corresponding 
post-measurement state. Allowing sufficient time for this remainder to 
reach a steady state, then moving to a new resonator, provides a method 
for measurements on individual mechanical modes.

The matrix coefficients for Eq.~(\ref{LyapRicc}) can be determined from Eqs.~(\ref{driftdiffusion}), (\ref{driftsystem}) and (\ref{Rdriftdiffusion})-(\ref{Rdriftdiffusion1}) above using the following Hamiltonians expressing the internal system coupling and the coupling between system and input modes:
\begin{align}
     H_s=&H_k\, ,\\
H_{C}=&\sqrt{\kappa}(XX_{\text{out}}+PP_{\text{out}})+\sqrt{\tau}(XX_{\text{out}}'+PP_{\text{out}}')\\&
+\sqrt{\gamma} \sum_j (X_jX_{j, \text{out}}'+P_jP_{j, \text{out}}')\,,\nonumber
\end{align}
where all the $X$ and $P$ are conjugate operators, the 
subscript `out' describes the output fields, and the primed quadratures are unmonitored output modes. The
unmonitored cavity decay represents genuine photon losses (e.g., 
scattering or absorption of the cavity) and is characterised by a rate $\tau$. The system also dissipates interacting with the thermal environment of each mechanical mode. We assume for simplicity that each mechanical resonator is locally in contact with its own purely dissipative environment, each of which is characterised by the same damping rate $\gamma$ and temperature $T$. The state of each mechanical bath is a Gaussian thermal state characterized by the covariance matrix $\sigma_{j, \text th}=(n_j +\frac12) \mathbb{I}$, with $n_j=\left[\exp \left(\frac{\hbar \Omega_j }{k_B T}\right)-1\right]^{-1}$.
\section{Continuous Monitoring for Gaussian Transformations}\label{sec:continuous-monitoring}

In this Section, we will demonstrate that continuous monitoring of the cluster state can reproduce, in certain regimes, the same results of a Gaussian transformation implemented directly on a cluster through projective measurements. The key Gaussian operations that we are going to investigate are the identity $\mathbb{I}\rightarrow\{0,0,0,0\}$, the Fourier transform $F\rightarrow\{1,1,1,0\}$, the shearing operation $S(1)\rightarrow\{1,0,0,0\}$ and the two-mode operation $CZ$. As recalled in Sec~\ref{sec:MBQC}, the decomposition in \refeq{symplectic_decomposition} links a set of four numbers $\{\lambda_1,\dots,\lambda_4\}$ to the associated Gaussian transformation induced by the four measurements $p+\lambda_jq$ performed on the first four nodes of a five-mode linear cluster ($j=1,\dots,4$). Via \refeq{X_phi}, each of these measurements is related to the choice of the driving phase given by 
\begin{equation}
	\phi_j=\arctan\frac{1}{\lambda_j}\ .
\end{equation}
As said, these operations are (for arbitrary shearing parameters) universal for multimode Gaussian transformations. In particular, the identity operation covers displacements in the decomposition of Gaussian transformations discussed in Sec~\ref{sec:MBQC}.

Before proceeding let us stress two observations regarding the initial state of the system and the figure of merit that we use. First, let us assume that we want to perform a single-mode operation on a generic input state. The standard procedure to accomplish this in a MBQC setting is to attach the latter to a four-node cluster via a $CZ$ operation. Without loss of generality, we take the input state to be a  vacuum state squeezed in momentum to the same degree as the constituents of the ancillary cluster. This makes the system before the first measurement equivalent to a five node linear cluster, which could in turn be prepared following the protocol described in Ref.~\cite{houhou2015generation}. The preparation scheme ends completely before detection takes place, and so the drive fields for generation and measurement do not overlap. This will be the initial state of the system that we will consider (see below for two-mode operations). Second, let us recall that the objective of this analysis is to demonstrate that the continuous monitoring scheme is functionally equivalent to the direct projective measurements required by standard MBQC. To this end, the states whose fidelity we calculate are those of the outputs from continuous monitoring and from projective measurements on a cluster state. Therefore, a high fidelity indicates that continuous monitoring well approximates projective measurements. This is not the same as saying that the output state has a high fidelity with the expected outcome of a computation. For this to occur, a high level of squeezing is known to be required in general \cite{menicucci2006universal} --- with the limit for fault-tolerant computation currently set at approximately $20\dB$ \cite{menicucci2014fault}. 

The set of parameters that describe the opto-mechanical system under consideration is relatively large, and is given in \reftab{table}. This includes: the linearized interaction strength $\alpha g$, the mechanical dissipation rate $\gamma$, the mechanical bath temperature $T$, the cavity decay rate $\kappa$, the unmonitored losses $\tau$, the detector efficiency $\eta$, the squeezing of the post-measurement state of the homodyned mode $r_\text{post-meas}$ and the squeezing of the cluster $r_\text{cluster}$ \footnote{In our numerical analysis, the squeezing of the cluster $r_\text{cluster}$ is fixed at $3\dB$, which is the theoretical limit for standard parametric squeezing techniques. Note however that this limit has very recently been overcome using reservoir engineering \cite{lei2016quantum}.}. In addition, one may also tune the monitoring time $t_\text{mon}$ for each step in the measurement process.

Given the amount of parameters involved, it is of relevance to individuate which of them are going to be determinant for the purposes of MBQC. To this end, we are now going to provide simulations of the continuous monitoring procedure varying certain parameters and progressively eliminating from consideration those with the weakest contribution. Hence, this analysis will not only establish the possibility to perform MBQC using continuous monitoring rather then direct measurements, but it can also be seen as instrumental at a quantitative level for setting benchmarks and driving experimental efforts in using opto-mechanics for advanced quantum information tasks.  

\begin{table}
	\begin{tabular}{|p{2cm}||p{2.6cm}|p{2.6cm}|}
 		\hline
		Parameter & Set 1 & Set 2 \\
		& (realistic) & (close to ideal) \\
		\hline
		$\eta$ & $0.99$ & $1$\\
		$\frac{\gamma}{2\pi}$ & $8\Hz$ & $0\Hz$\\
		$\frac{\kappa}{2\pi}$ & $0.33\MHz$ & $0.1\MHz$\\
		$\tau$ & $0.01\kappa$ & $0$\\
		$\alpha g$ & $0.35\MHz$ & $0.35\MHz$\\
		T & $1$~mK & $0$~K\\
		$r_{\text{post-meas}}$ & $10\dB$ & $20\dB$\\
		$r_{\text{cluster}}$ & $3\dB$ & $3\dB$\\
		\hline
	\end{tabular}
	\caption{Experimentally-motivated (Set 1) and close-to-ideal (Set 2) values for parameters used in the simulations. Where a parameter is varied across a simulation (see Figs.\ref{irrelevant}-\ref{Contour}), that value takes precedence over the table value. The resonators have frequencies $2\pi j~11\MHz$ ($j=1,\dots,N$).}\label{table}
\end{table}

\begin{figure}
	\includegraphics[width=\columnwidth]{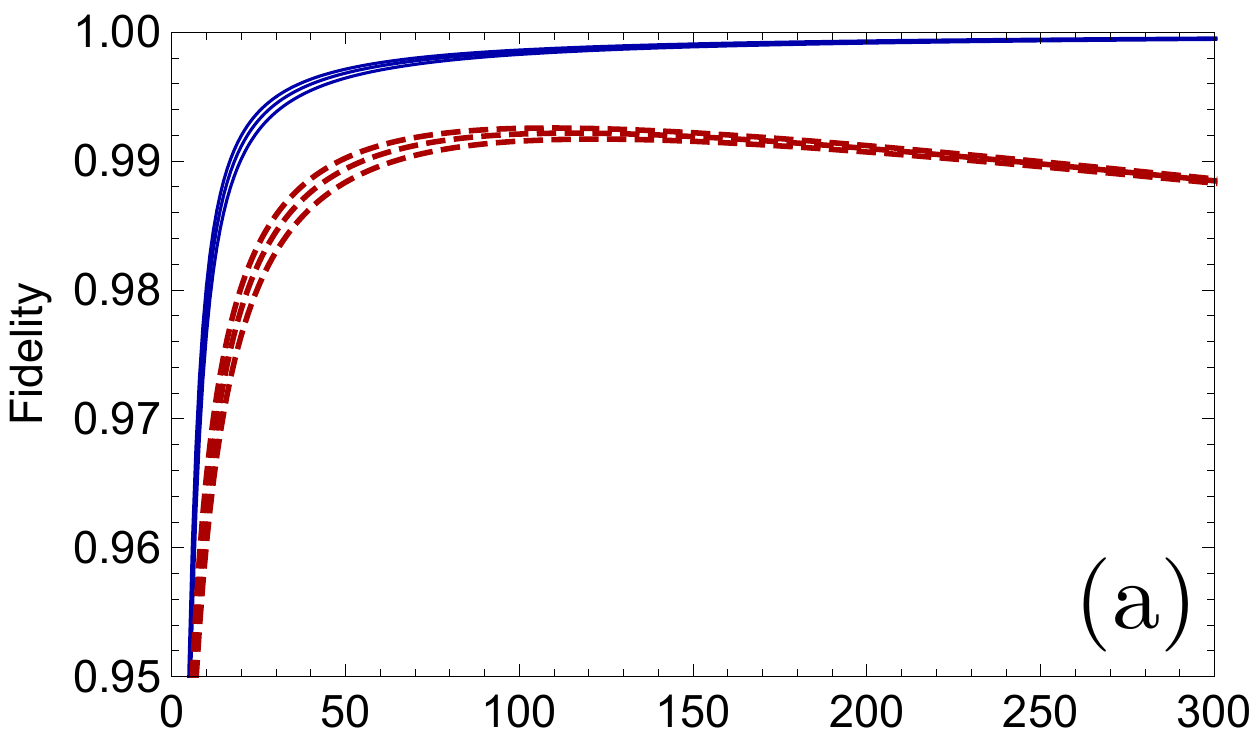}
	\includegraphics[width=\columnwidth]{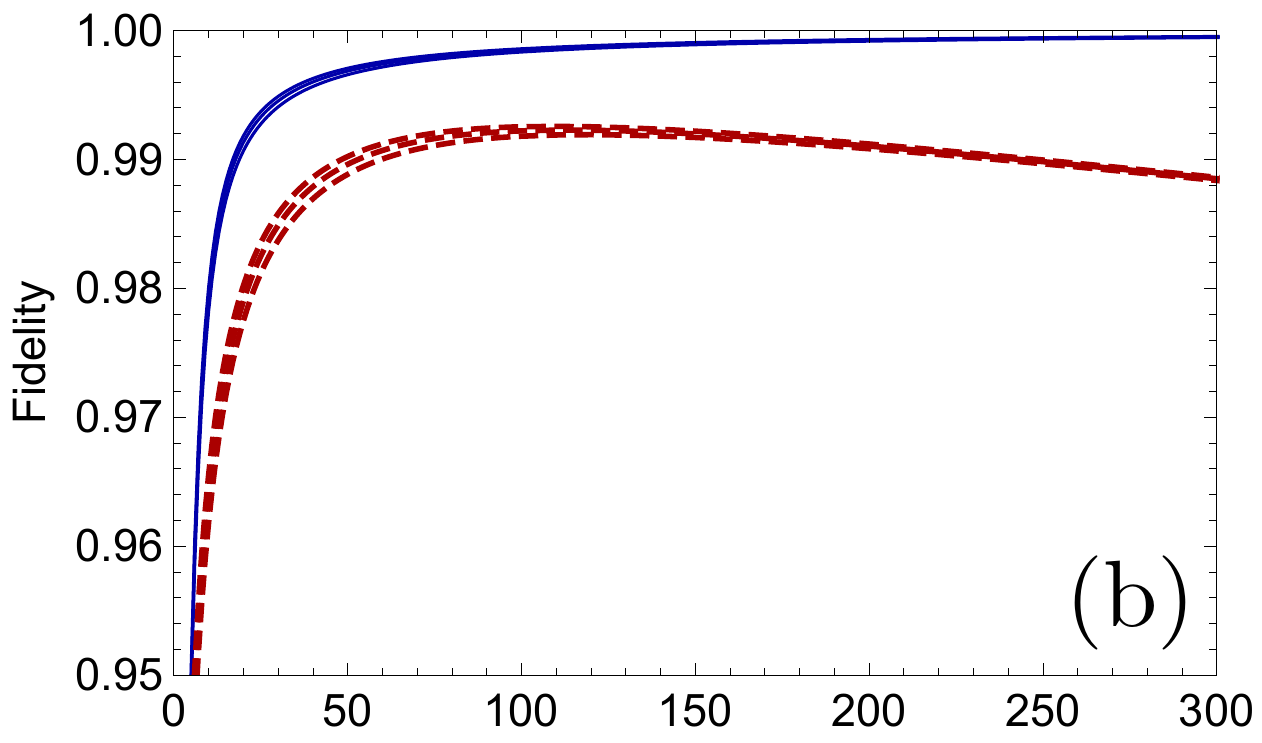}
	\includegraphics[width=\columnwidth]{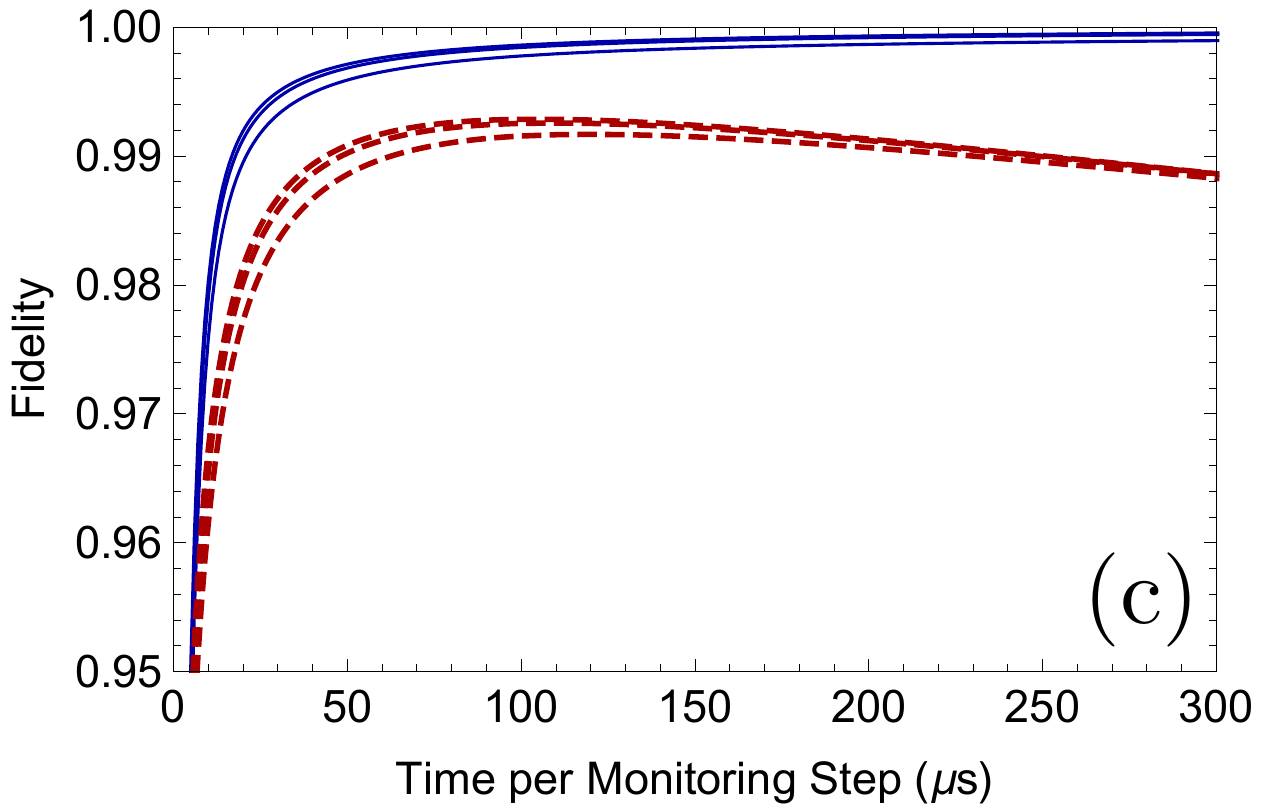}
	\caption{From top to bottom, effect of the variation of the detector efficiency $\eta$,  unmonitored losses $\tau$, and squeezing of the post-measurement state of the homodyned mode $r_{\text{post-meas}}$ on the highest achievable fidelity on application of the gate $S(1)$ to a 5-node linear cluster against the monitoring time per step. Except for the noise parameter under consideration all the parameters are set as per \reftab{table} with solid and dashed lines denoting close-to-ideal (Set~2) and realistic (Set~1) parameters respectively. In each plot the curves corresponding to each set of parameters bunch together (particularly for long monitoring times). From top to bottom of each bunch the parameter under consideration is varied: (a) the detector efficiency $\eta=1, 0.9, 0.8$ (b) the unmonitored losses $\tau=0.01\kappa, 0.05\kappa, 0.1\kappa$ and (c) the squeezing of post-measurement state of the homodyned mode $r_{\text{post-meas}}=20, 10, 5$ dB. These plots show that these three noise mechanisms are only minimally detrimental for successfully implementing Gaussian transformations on a mechanical cluster state, and that their effect is more and more negligible for long monitoring times. 
	} \label{irrelevant}
\end{figure}

To show the effects of this collection of parameters, we vary each of them individually within two sets (see \reftab{table}):\textit{ (i)} Set~1 refers to realistic values of the parameters involved, as guided by recent experimental achievements \cite{lei2016quantum}; \textit{(ii)} Set~2 refers to close-to-ideal settings in which all losses are neglected, near perfect homodyne measurements are achievable, and access to the sideband resolved regime is deep. In the majority of the analysis below, our target is to apply the gate given by the shearing operation $S(1)$, however we will also demonstrate that gate choice has little effect on the final fidelity.

\begin{figure}
	\includegraphics[width=\columnwidth]{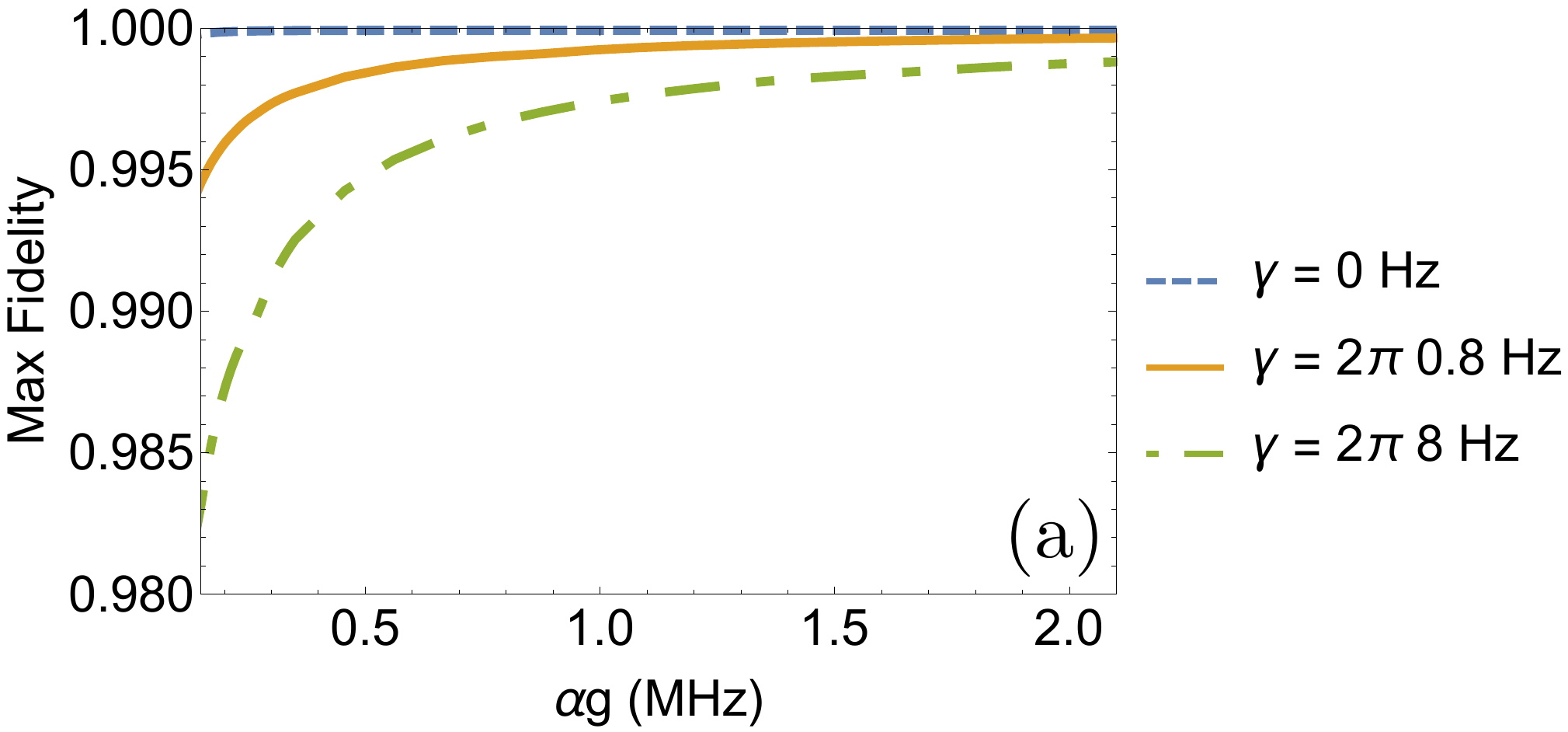}
	\includegraphics[width=\columnwidth]{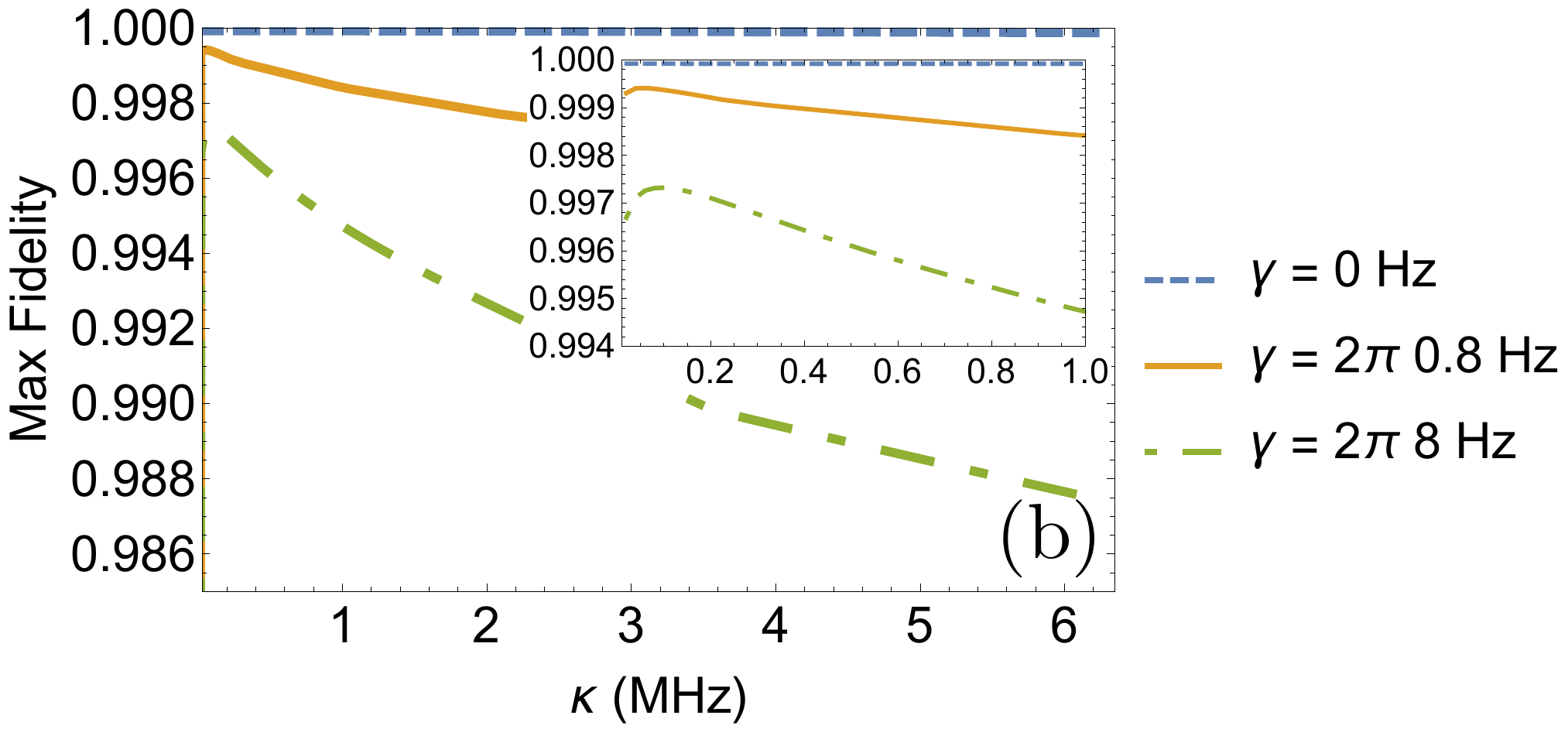}
	\caption{Variation of (a) the interaction strength $\alpha g$ and (b) the cavity decay rate $\kappa$. These plots show the maximum fidelity achieved over long monitoring times using realistic parameters as per Set~1 of \reftab{table} (except for the mechanical losses $\gamma$ whose value is indicated in the legend).}
	\label{tunable}
\end{figure}

We start our analysis by considering the role played by the noise parameters $\eta$, $\tau$, and $r_\text{post-meas}$. The plots in Fig.~\ref{irrelevant} show that their effect on the highest achievable fidelity is relatively minimal both in the close-to-ideal and realistic settings. This is due to the fact that the monitoring scheme continuously gathers information on the mechanical state, driving it towards an eigenstate associated with $X_{\phi}$. This suggests that small losses --- due either to genuine photon loss from the cavity ($\tau$) or photons missed at the detector end ($\eta$) --- are ineffectual at disturbing the measurement outcome, provided enough monitoring time is given. In other words, small amounts of measurement losses can be compensated for by increasing the time for which the system is monitored. For similar reasons, the deviations from ideal homodyne measurements due to a finite degree of squeezing of the post-measurement state have little effect. However note that we maintain $r_\text{cluster}<r_\text{post-meas}$ --- a condition that is typically met in experimental settings. In general, all the plots of Fig.~\ref{irrelevant} show that when the remaining parameters are set in close-to-ideal conditions (Set~2), unit fidelity is in fact achieved for long monitoring time. When realistic parameters (Set~1) are considered, obviously unit fidelity  cannot be achieved any more (as the mechanical resonators tend to thermalize asymptotically), however it is still true that the effect of variations in $\eta$, $\tau$, and $r_\text{post-meas}$ is negligible for large monitoring times.  Notice also that, for all the curves that refer to realistic parameters (dashed curves), the fidelity achieves a maximum for a finite monitoring time $t_\text{mon}$. As it is reasonable to expect when mechanical losses are present, this indicates that one can optimize $t_\text{mon}$, an opportunity that we will address later on.

Let us now turn our attention to the cavity output rate $\kappa$ and the effective coupling strength $\alpha g$. As detailed in \refsec{sec:optomechanical-setting}, our analysis is valid in the side-band resolved regime and when the rotating wave approximation holds, respectively when $\kappa$ and $\alpha g$ are both much lower than the mechanical frequencies. However, as long as these conditions are met, there is still room for optimizing their values. Regarding $\alpha g$, it determines the strength of the effective measurement that we are performing on the mechanical resonators by monitoring the cavity output. Hence, in principle, the larger it is the more effective our measurement strategy is. The upper panel of \reffig{tunable} shows that this is indeed the case. However, depending on the values of the remaining parameters, it might not be necessary to increase $\alpha g$ up to the limit of validity of the rotating wave approximation, since a fidelity close to one can be achieved for relatively weak coupling. That is, once $\alpha g$ is sufficiently large it ceases to be beneficial to increase it further.

\begin{figure}
	\includegraphics[width=\columnwidth]{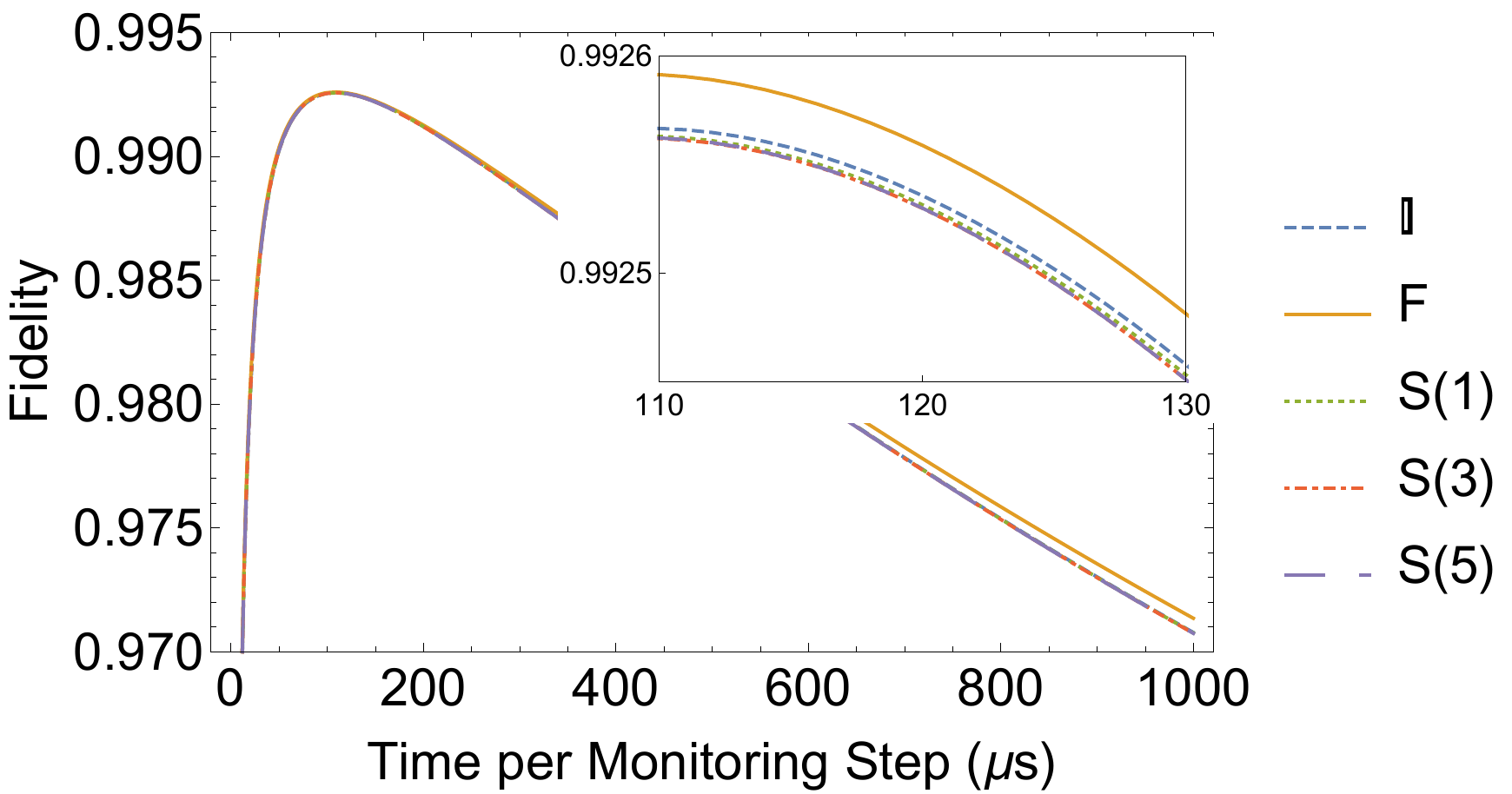}
	\caption{Applying the operations of identity $\mathbb{I}$, Fourier transform $F$, and shearing $S(\lambda)$ with $\lambda=1,3,5$ under the realistic parameters of \reftab{table}. The final fidelity of the output of these operations is plotted against the monitoring time for each step.}
	\label{Gates}
\end{figure} 

Regarding the cavity output rate $\kappa$, one expects that the deeper we are in the side-band resolved regime the better --- as long as the monitoring time is large enough so that the measurement of the (small portion of) radiation emerging from the cavity is still effective. This is in fact shown in \reffig{tunable}(b), where it is also evident however that for non-zero mechanical losses $\gamma$ a trade-off appears. This is due to the fact that for very small $\kappa$ the monitoring time required to effectively measure the mechanical resonators starts to be too large and the mechanical losses spoil the MBQC-induced dynamics.

\begin{figure}
	\includegraphics[width=\columnwidth]{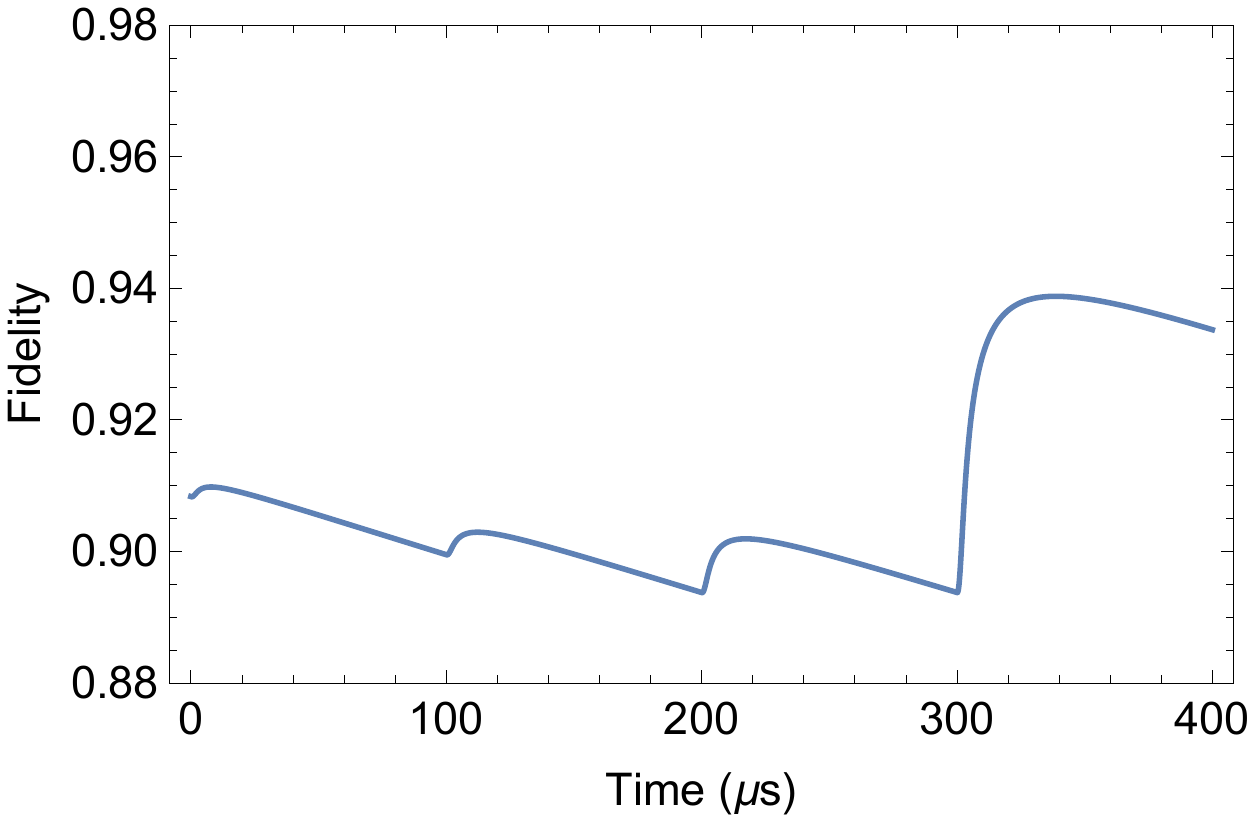}
	\includegraphics[width=\columnwidth]{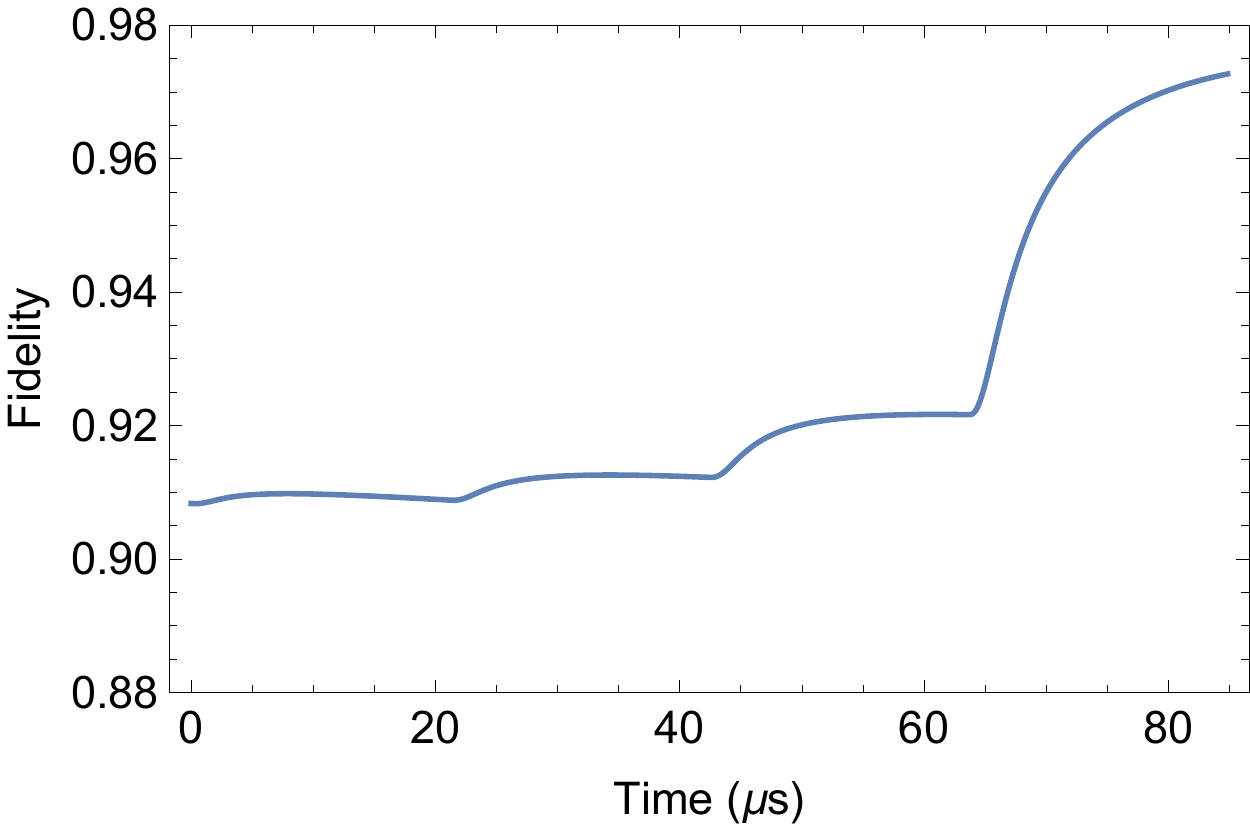}
	\caption{The fidelity of the identity operation with equally spaced step intervals (left) and with optimised steps (right). Without optimisation there are long periods where the fidelity is not increased by the monitoring scheme, and in fact is actively decreased by exposure to the mechanical damping. With optimisation, the ceiling for the fidelity is raised and the time required to reach high fidelities is significantly lessened. These simulations also consider realistic parameters (Set~1), with the exception of $T=10$~K, emphasising the effect of the optimised monitoring steps. The optimised steps for this scenario are $21.4,$ $21.2,$ $21.1$ and $21.1~ \mu$s.}
	\label{Optima}
\end{figure}

We next demonstrate that the choice of the gate to be implemented has little effect on the final fidelity, as one would ideally require when performing a computation. Using the set of realistic parameters (Set~1) we observe that the three key single-mode gates [plus $S(3)$ and $S(5)$] achieve similar final fidelity curves, over the time interval allotted per measurement step (Fig.~\ref{Gates}). This shows that the continuous monitoring scheme is not greatly affected by the choice of measurement to be performed and that a computation need not be optimised against using certain operations.

\begin{figure*}[t!]
	\includegraphics[width=\textwidth]{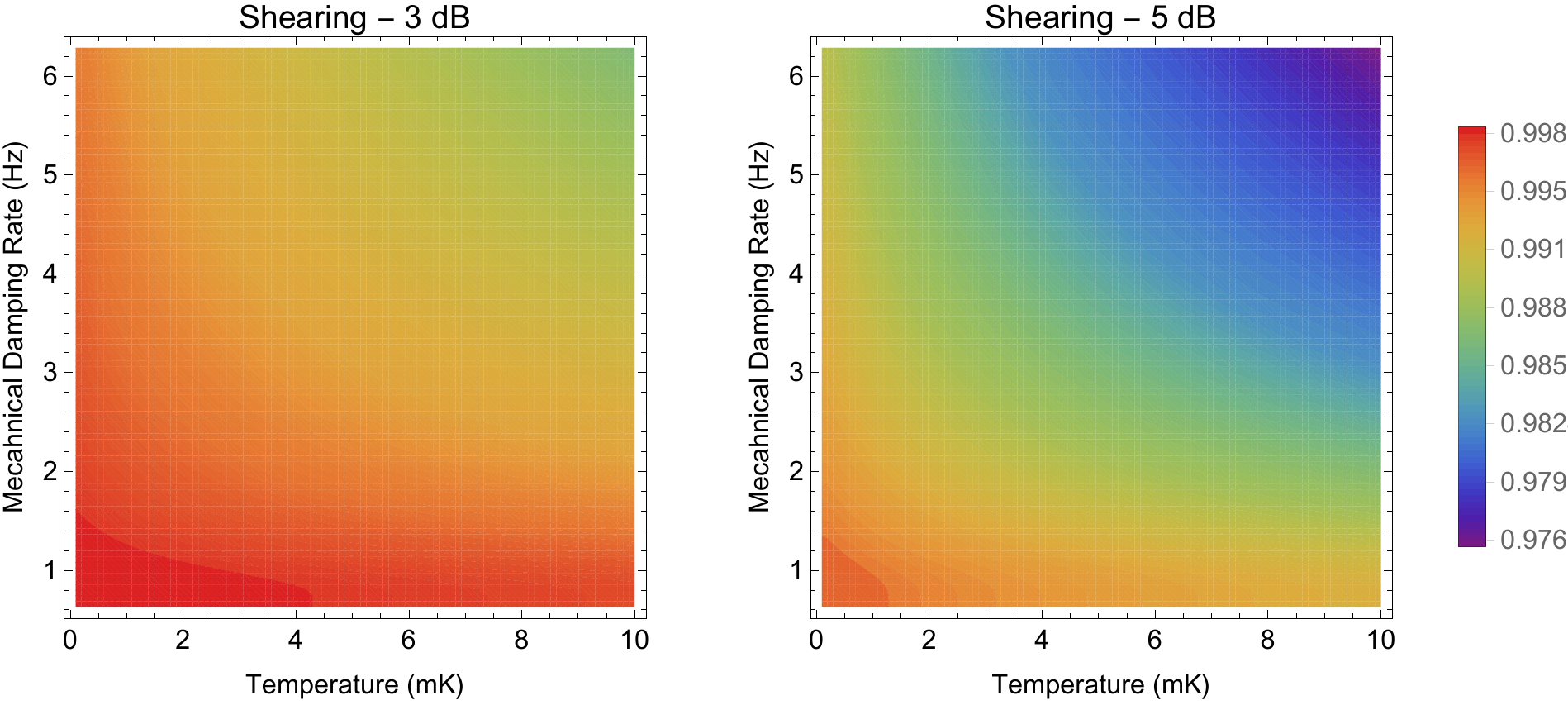}
	\includegraphics[width=\textwidth]{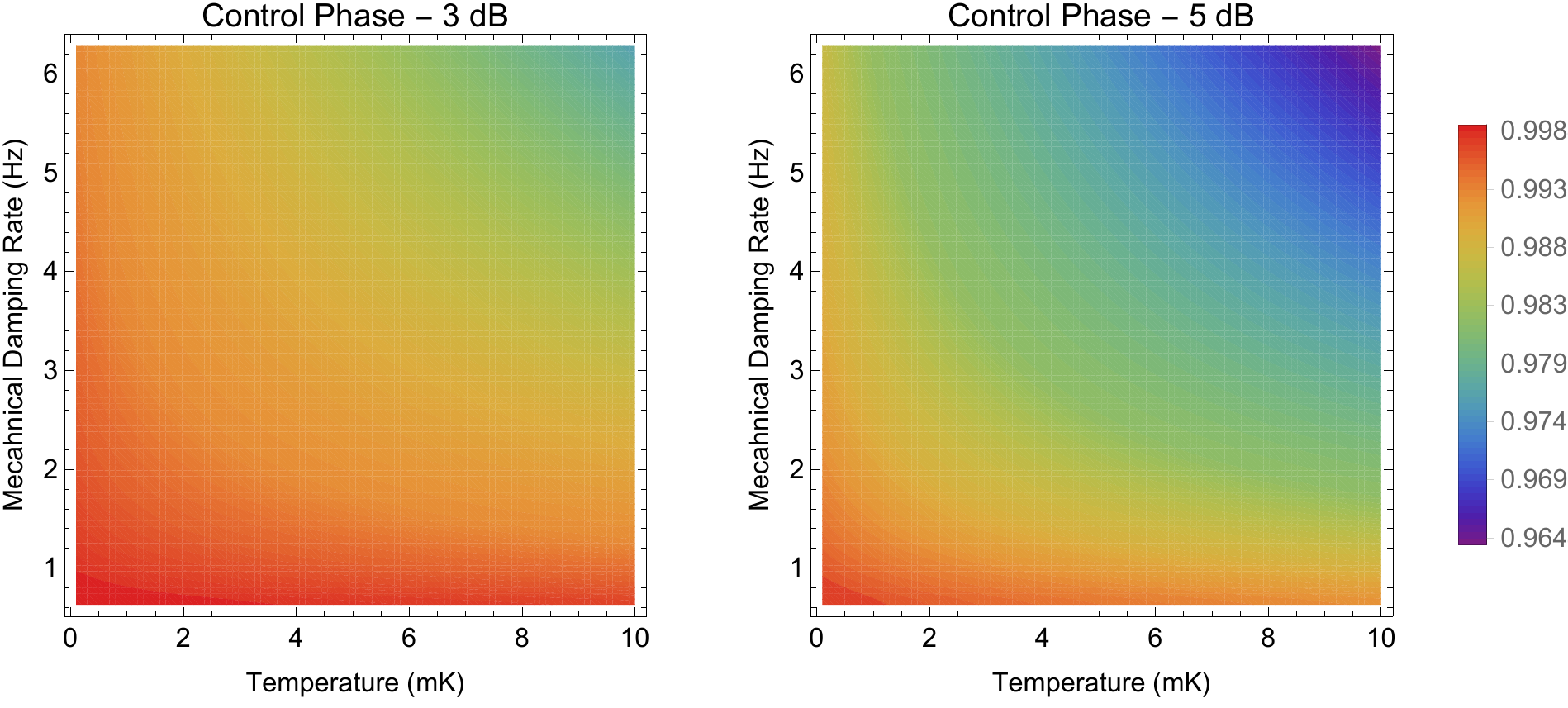}
	\caption{The maximum achievable fidelity with optimised monitoring intervals for the continuous monitoring implementation of the shearing $S(1)$ and $CZ$ operations. Increased squeezing increases the susceptibility of the output to damage from the thermal environment.}
	\label{Contour}
\end{figure*}

We have noticed above (see both Figs.~\ref{irrelevant} and \ref{Gates}) that the monitoring time per step can be optimized. Now we may demonstrate more quantitatively the effect of the length of the monitoring interval, the time $t_\text{mon}$ taken for a single measurement step. In all analyses thus far the monitoring has been performed with each step taking an equal fraction of the total time for the scheme to complete. However, the monitoring interval is an easily tunable parameter and heuristically has large effects on the quality of the computation. This is because short intervals do not allow the measured node to reach a state with an acceptable degree of squeezing, whereas long intervals allow mechanical damping to dominate the dynamics. We can optimise by making the length of each monitoring step independent of the others such that the fidelity never decreases. In \reffig{Optima} we show that optimising the monitoring interval has the beneficial effect of increasing monotonically with time the fidelity of the output mode with the target state. Physically we want the mechanical system to spend minimal time in the fire and consequently to be less affected by its environment. Notice that the effects of this optimisation are to simultaneously raise the ceiling on the achievable fidelity and significantly decrease the length of the monitoring process.

After the analysis performed so far, only $T$, $\gamma$ and $r_\text{cluster}$ remain from \reftab{table}. Since these appear to be the most relevant parameters we provide a more exhaustive analysis of their effects, both for the key single-mode operations and for the two-mode $CZ$ operation. Recall from \refsec{sec:MBQC} that  for the $CZ$ operation a four node cluster is topologically equivalent to a minimal dual rail configuration [\reffig{Cluster}~(c)]. It is the latter that we are going to use for the $CZ$ operation.

Using the realistic parameters (Set~1), \reffig{Contour} shows the maximum fidelity achieved with optimised monitoring intervals as the temperature and mechanical dissipation are varied over the shearing $S(1)$ and $CZ$ operations, for $r_\text{cluster}=3\dB$ and $5\dB$. We can clearly see that high levels of fidelity, above 95\%, can be achieved, in all the cases considered, for temperature $T$ and mechanical damping rate  $\gamma$ within reach of current technology. This establishes opto-mechanical systems as a promising candidate for advanced computational tasks. As expected, increasing $\gamma$ or $T$ has deleterious effects on the fidelity. Additionally, increasing the squeezing parameter $r_\text{cluster}$ reduces the achievable fidelity for a particular pair $(\gamma, T)$. 
We expect that for larger squeezing values the region of the plots with large fidelity becomes more localised around the point $(\gamma,T)=(0,0)$.

\section{Conclusions}\label{sec:conclusion}

To conclude, we generalised an existing back-action evading measurement scheme in optomechanics  to  the  case  in  which  the  cavity  interacts  directly  with  a  collection  of  resonators. This allowed us to demonstrate the effectiveness of continuous monitoring in performing QND measurements on a mechanical cluster state. In particular, we showed that arbitrary multi-mode Gaussian transformations can be implemented with this method within the same experimental set-up that allows for the preparation \cite{houhou2015generation} and reconstruction \cite{moore2016Recon} of a mechanical cluster state. Therefore this work provides a significant step towards universal measurement-based computation with mechanical resonators.

We examined our procedure in a variety of conditions relevant to experiments, noting the prevalent impact of temperature and mechanical damping on the success of the computation, while showing the limited effect of a large set of other parameters. In general, low temperatures and mechanical dissipation rates are a fundamental requirement to replicate the results of projective measurements performed directly on a cluster state --- thus allowing for any computational step to occur. The requirements become more stringent when the degree of squeezing of the cluster is increased, a necessity for approaching fault tolerant computation.

In order to achieve universal computation a non-Gaussian operation has to be added to the set of operations considered here. This will necessitate involving a non-Gaussian element in the network, whether taking  advantage of a non-Gaussian measurement, a non-linear optomechanical dynamics,  or an already prepared non-Gaussian resource. Promisingly in this direction, it has been shown that opto- and electro-mechanical systems can intrinsically host non-linearities in various settings \cite{sankey2010strong, dykman2012fluctuating, rips2014nonlinear, fonseca2016nonlinear}, and this possibility has been suggested for engineering non-Gaussian states, dynamics, and measurements \cite{rips2012steady, akram2013entangled, borkje2013signatures, montenegro2014nonlinearity, filip2015transfer, brunelli2015out, teklu2015nonlinearity, albarelli2016nonlinearity, hoff2016measurement}. The latter could potentially be exploited to unlock the universality of computation and this shall be the topic of future work.

\acknowledgements

The authors thank M. Genoni, A. Nunnenkamp, M. Paternostro and A. Xuereb for useful discussions. AF and OH acknowledge funding from the EPSRC project EP/P00282X/1. DM acknowledges funding from the EPSRC.

	\bibliography{references}

\end{document}